\documentclass[twocolumn]{cccg}
\usepackage{graphicx}
\usepackage{url}
\usepackage{latexsym}

\newcommand{\squeezelist}{\setlength{\itemsep}{0pt}}
\newenvironment{pf}{\unskip{\bf Proof:}}{\unskip{\hfill $\Box$}}

\newcommand{\lemlab}[1]{\label{lemma:#1}}
\newcommand{\theolab}[1]{\label{theo:#1}}

\newcommand{\tablab}[1]{\label{tab:#1}}
\newcommand{\figlab}[1]{\label{fig:#1}}
\newcommand{\seclab}[1]{\label{section:#1}}

\newcommand{\lemref}[1]{\ref{lemma:#1}}

\newcommand{\theoref}[1]{\ref{theo:#1}}

\newcommand{\tabref}[1]{\ref{tab:#1}}
\newcommand{\figref}[1]{\ref{fig:#1}}
\newcommand{\eqref}[1]{(\ref{eq:#1})}
\newcommand{\secref}[1]{\ref{section:#1}}

\newtheorem{theorem}{Theorem}
\newtheorem{lemma}{Lemma}

{\catcode`\@=11
\gdef\setft#1#2#3{%
\def\@oddfoot{
{\setbox0=\hbox{#1}
\setbox1=\hbox{#3}
\ifdim\wd0>\wd1
\dimen0=\wd0
\box0\hfil#2\hfil\hbox to\dimen0{\hfil\hfil\box1}
\else \dimen0=\wd1
\hbox to\dimen0{\box0\hfil }\hfil#2\hfil\box1 \fi
}}} }

\def\complaint#1{}
\def\withcomplaints{
\newcounter{mycomplaints}
\def\complaint##1{\refstepcounter{mycomplaints}%
\ifhmode%
\unskip%
{\dimen1=\baselineskip \divide\dimen1 by 2 %
\raise\dimen1\llap{\tiny -\themycomplaints-}}\fi%
\marginpar{\tiny [\themycomplaints]: ##1}}%
}

\renewenvironment{pf}{
\noindent\textbf{Proof:}}{\unskip{\hfill $\Box$}}

\let\oldendpf=\endpf
\def\endpf{\oldendpf\par\medskip}

\title{\bf PushPush and Push-1 are NP-hard in 2D}
\author{%
Erik D. Demaine 
\thanks{
Dept.\ Comput\ Sci., Univ.\ Waterloo,
Waterloo, Ontario N2L 3G1, Canada.
\texttt{\{eddemaine, mldemaine\}@\penalty \exhyphenpenalty uwaterloo.ca}.
}
\and Martin L. Demaine\footnotemark[1]
\and
Joseph~O'Rourke\thanks{
Dept.\ Comput.\ Sci., Smith Col\-lege, North\-ampton,
MA 01063, USA.
\texttt{orourke@\penalty \exhyphenpenalty cs.smith.edu}.
Supported by NSF grant CCR-9731804.}
}
\date{}

\begin{document}
\maketitle
\begin{abstract}
We prove that two pushing-blocks puzzles are intractable in 2D.
One of our constructions improves
an earlier result that established intractability in
3D~\cite{ppnph-os-99} for
a puzzle inspired by the game {\em PushPush}.
The second construction answers a question we
raised in~\cite{ddo-ppnph2d-00} for a variant
we call {\em Push-1}.
Both puzzles
consist of unit square blocks on an integer lattice;
all blocks are movable.
An agent may push blocks (but never pull them) in attempting
to move between given start and goal positions.
In the PushPush version, the agent can only push one block at
a time, and moreover when a block is pushed it slides the
maximal extent of its free range.
In the Push-1 version, the agent can only push one block
one square at a time, the minimal extent---one square.
Both NP-hardness proofs are by reduction from SAT,
and rely on a common construction.
\end{abstract}

\begin{figure*}[htbp]
\centering
\includegraphics[width=0.475\textwidth]{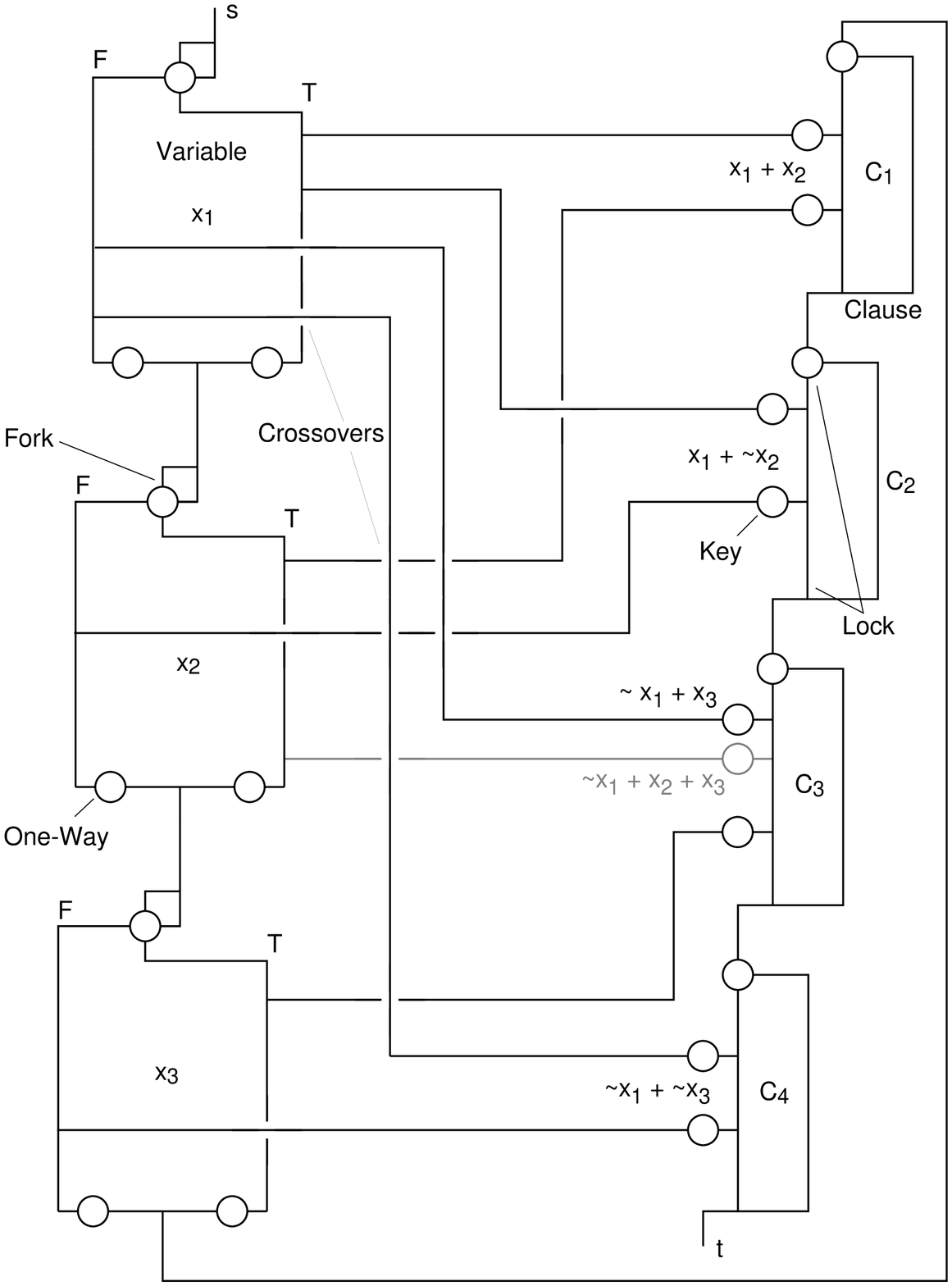}
\hfill
\includegraphics[width=0.475\textwidth]{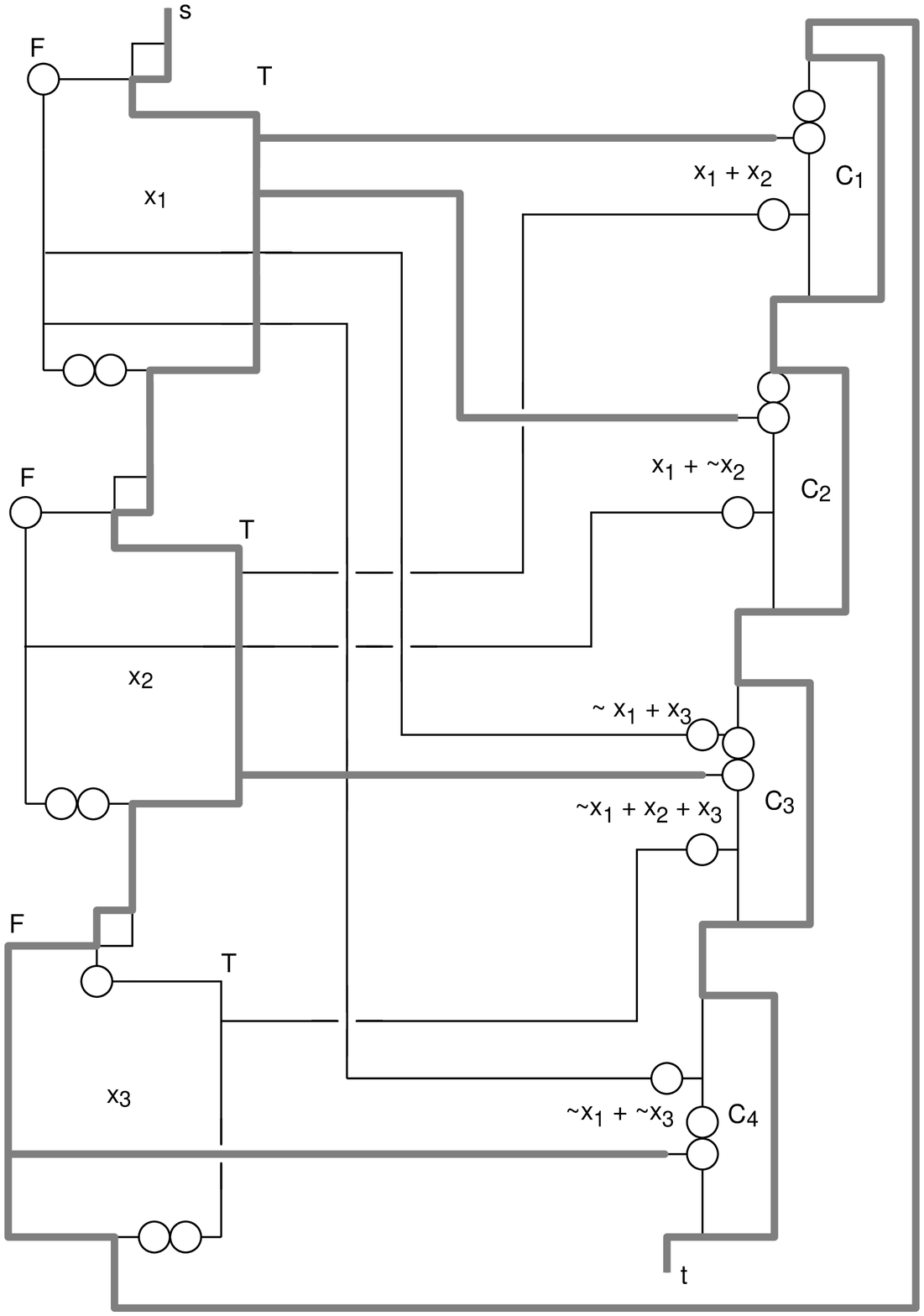}
\caption{Left: Complete construction of the NP-hardness reduction for
              PushPush from \protect\cite{ppnph-os-99} for the formulas in
              Eq.~(1) and Eq.~(2)
  (including the shaded portion).  Right: Solution path for Eq.~(2).
[Based on Figs.~6 and~7 in~\protect\cite{ppnph-os-99}.]
}
\figlab{3D.SAT}
\end{figure*}

\section{Introduction}
\seclab{Introduction}
There are a variety of ``sliding blocks'' puzzles whose
time complexity has been analyzed.
One class, typified by the 15-puzzle so heavily
studied in AI, permits an outside
agent to move the blocks.
Another class falls more under the guise of
motion planning.
Here a robot or internal agent plans a
path in the presence of movable obstacles.
This line was initiated by a paper of Wilfong~\cite{w-mppmo-91},
who proved NP-hardness of a particular version in which the
robot could pull as well as push the obstacles, which were
not restricted to be squares.
Subsequent work sharpened the class of problems by weakening
the robot to only push obstacles, and by 
restricting all obstacles to be unit squares.
Even this version is NP-hard when some blocks may be
fixed to the board (made unpushable)~\cite{do-mpams-92}.

One theme in this research has been to establish stronger degrees
of intractability, in particular,   
to distinguish between NP-hardness and PSPACE-completeness, 
the latter being the stronger claim.  
The NP-hardness proved in~\cite{do-mpams-92} was 
strengthened to PSPACE\--completeness in an unfinished 
man\-u\-script \cite{bos-mpams-94}.  
More firm are the results on
Sokoban, a
computer game that restricts the pushing robot to only push one block at
a time, and requires the storing of (some or all) blocks into 
designated ``storage
locations.''
This game was proved NP-hard in~\cite{dz-sompp-95},
and PSPACE-complete by Culberson~\cite{c-spc-99}.  

Here we emphasize another theme: finding a nontrivial version of
the game that is {\em not\/} intractable.  To date only the most uninteresting
versions are known to be solvable in polynomial time, for example,
where the robot's path must be monotonic~\cite{do-mpams-92}.
To explore the variety of pushing-block puzzles 
it is useful classify them according to these characteristics:
\begin{enumerate}
\squeezelist
\item Can the robot pull as well as push?
\item Are all blocks unit squares, or may they have different shapes?
\item Are all blocks movable, or are some fixed to the board?
\item Can the robot push more than one block at a time?
\item Is the goal for the robot to move from $s$ to $t$,
or is the goal for the robot to push blocks into storage locations?
\item The dimension of the puzzle: 2D or 3D?
\item Do blocks move the minimal amount, exactly how far they
are pushed, or do they slide the maximal amount of their
free range?
\end{enumerate}

If our goal is to find the weakest robot and most
unconstrained puzzle conditions that still lead to intractability, 
it is reasonable to consider robots who can only push~(1),
and to restrict all blocks to be unit squares~(2), 
as in \cite{do-mpams-92,dz-sompp-95,c-spc-99}, for
permitting robots to pull, and permitting blocks of other shapes,
makes it relatively easy to construct intractable puzzles.
It also makes sense to explore the goal of simply finding a path~(5)
as in \cite{w-mppmo-91,do-mpams-92}, rather than
the more challenging task of
storing the blocks as in Sokoban~\cite{dz-sompp-95,c-spc-99}.
Allowing the robot to move in 3D~\cite{ppnph-os-99} 
gives it more ``power'' than
it has in 2D~\cite{ddo-ppnph2d-00}~(6), so we focus on 2D.


The versions explored in this paper superficially seem that they might
lend themselves to polynomial-time algorithms: in both,
the robot can only push
one block~(4), and all blocks are pushable~(3).
We explore two different versions,
the first again inspired by a computer game, {\em PushPush\/}.\footnote{
        The earliest reference we can find to the game
        is a version written for the Macintosh by
        Alan Rogers and C.M. Mead III, Copyright 1994,
        \url{http://www.kidsdomain.com/down/mac/pushpush.html}.
        Another version for the Amiga was written by
        Luigi Recanatese in 1997,
        \url{http://de.aminet.net/aminet/dirs/game_think.html}.
	See also \url{http://daisy.uwaterloo.ca/~eddemain/pushingblocks/}
        for our implementation.
}  
The key difference in this game is in characteristic~(7):
when a block is pushed, it necessarily slides 
(as without friction) the maximal
extent 
of the available empty space in the direction in which it was
shoved. 
It was established in~\cite{ppnph-os-99} that the problem is
intractable in 3D, but its status in 2D was left open in that paper.
Here we settle the issue by extending the reduction to 2D.

Continuing the theme of weakening the robot's capabilities,
we also study a version we call {\em Push-1},
with the same characteristics as PushPush except 
that the one pushed block
moves the minimal amount, just one square at a time.

Although our original proof for the hardness of
PushPush~\cite{ddo-ppnph2d-00} very much relied on maximal sliding,
the proof we offer in this paper establishes both games NP-hard
via the exact same construction.  We arrange so little freedom
that maximal and minimal sliding become the same.
We start in Section~\secref{PushPush} 
with the 3D PushPush construction from~\cite{ppnph-os-99},
whose overall structure
is followed in the new proof, described in Section~\secref{Push1}.
A summary of related results is presented in the
final section.

\section{PushPush in 3D}
\seclab{PushPush}
We first review the hardness proof from~\cite{ppnph-os-99},
which forms a skeleton for our proofs.
Observe that any $2 \times 2$ cluster of 
movable blocks is forever frozen to a PushPush or Push-1 robot, for there
is no way to chip away at this unit.  This makes it easy to
construct ``corridors'' surrounded by fixed regions to guide
the robot's activities.  
To describe the PushPush 3D construction, we use
an orthogonal graph, whose edges represent the corridors, understood
to be surrounded by sufficiently many $2 \times 2$ clusters to render any
movement outside the graph impossible.  The few movable blocks are
represented by circles.

\subsection{3D SAT Reduction}
\seclab{SAT.Reduction}
The reduction is from SAT, i.e., satisfiability of formulas in
conjunctive normal form.
The basic idea is to have variable ``gadgets'' or ``units'' 
that force the robot
to make a choice between two paths (setting the variable $x_i$ to {\sc t} or
{\sc f}).
Each variable gadget connects to the relevant clause gadgets.
The variable units are arranged in a linked chain that must be
visited in order, after which the clause units must visited
one after the other.
The clause units are impassable unless they were earlier
visited from a variable unit.
The only paths from $s$ to $t$ force 
the robot
to traverse all variables and then all clauses; 
so all clauses must be satisfied.

The complete construction for four clauses
$C_1 \wedge C_2  \wedge C_3 \wedge C_4$
is shown in 
Fig.~\figref{3D.SAT}, left.
Two versions of the clauses are shown in the figure:
an unsatisfiable formula (the dark lines),
and a satisfiable formula (including the shaded $x_2$ wire):
\begin{small}
\let\oldsim=\sim
\def\sim{\oldsim\!}
\begin{equation}
(x_1 \vee x_2) \wedge  (x_1 \vee \sim x_2) \wedge  (\sim x_1 \vee x_3) \wedge
(\sim x_1 \vee \sim x_3)
\end{equation}
\begin{equation}
(x_1 \vee x_2) \wedge  (x_1 \vee \sim x_2) \wedge  (\sim x_1 \vee x_2 \vee x_3) \wedge (\sim x_1 \vee \sim x_3)
\end{equation}
\end{small}%
Here we are using $\sim x$ to represent the negation of the variable $x$.
A path from $s$ to $t$ in the satisfiable version is illustrated in
Fig.~\figref{3D.SAT}, right.

Fig.~\figref{3D.SAT} identifies the essential components of the construction,
whose functionality will be duplicated under the more demanding Push-1
conditions:
\begin{enumerate}
\squeezelist
\item {\em Variable\/} units, where passage of the robot sets {\sc t} or
{\sc f}.
\item {\em Fork\/} gadgets, which force upon the robot
the variable-setting binary choice.
\item {\em One-Way\/} gadgets, which permit passage in one direction but
not the other.
\item {\em Clause\/} units, which may only be traversed if one of its
incident literals is {\sc t}.
\item {\em Lock \& Key\/} mechanisms, which prevent passage unless a
key block has been pushed.
\item {\em Crossover\/} units, which allow two ``wires'' to cross without
the possibility of leakage from one to the other.
\end{enumerate}
By far the greatest challenge is to construct 2D crossovers.

\section{Push-1 in 2D}
\seclab{Push1}

We concentrate on Push-1, and argue at the end our construction
also works for PushPush.  The complexity of the constructions
demand that we abandon the orthogonal graph representation, and
instead show all the blocks.  Fixed blocks (that is, effectively fixed blocks)
are shaded more darkly
than movable ones; the robot is depicted as a small disk.

\subsection{One-Way Gadget}
The simple {\em One-Way\/} gadget is shown in Fig.~\figref{One-Way}.
It only permits passage in the ``forward'' $a$-to-$b$ direction.
Note that after passage, it becomes a two-way corridor.
\begin{figure}[htbp]
\centering
\includegraphics[width=0.3\linewidth]{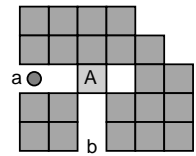}
\caption{Passage from $b$ to $a$ is prevented by block $A$.}
\figlab{One-Way}
\end{figure}

\subsection{Fork Gadget}
The {\em Fork\/} gadget, shown in
Fig.~\figref{Fork}, is the same mechanism
as employed in Fig.~\figref{3D.SAT}.

\begin{figure}[htbp]
\centering
\includegraphics[width=0.5\linewidth]{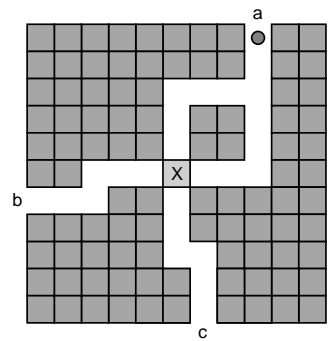}
\caption{Fork gadget. Block $X$ is initially at point $x$.}
\figlab{Fork}
\end{figure}
\begin{lemma}
A Fork gadget with central block $X$ in position $x$,
permits passage from $a$ to $b$, or from $a$ to $c$,
but once $b$ is reached from $a$, $c$ is inaccessible
via $x$; and symmetrically,
$b$ is inaccessible via $x$ once $c$ is reached from $a$.
\lemlab{Fork}
\end{lemma}
\begin{pf}
To reach $b$ from $a$, block $X$ must be pushed down into the corridor
heading toward $c$.  Then from $b$ it is no longer possible to
traverse that corridor from point $x$ toward $c$.
(Of course it might be possible to reach $c$ via some other route.)
\end{pf}

\subsection{Variable Unit}
It is now easy to construct a
a variable unit 
following the design in Fig.~\figref{3D.SAT}: a Fork upon
entrance, and a One-Way unit in the {\sc t} and in the {\sc f}
paths upon exit from the unit
(see Fig.~\figref{Clause}(left)).
The Fork prevents leakage into
the negated half by Lemma~\lemref{Fork}.
Note also the variable-clause wires have been spit into
one-way wires, for reasons to be explained shortly.

\begin{figure}
\centering
\includegraphics[width=\linewidth]{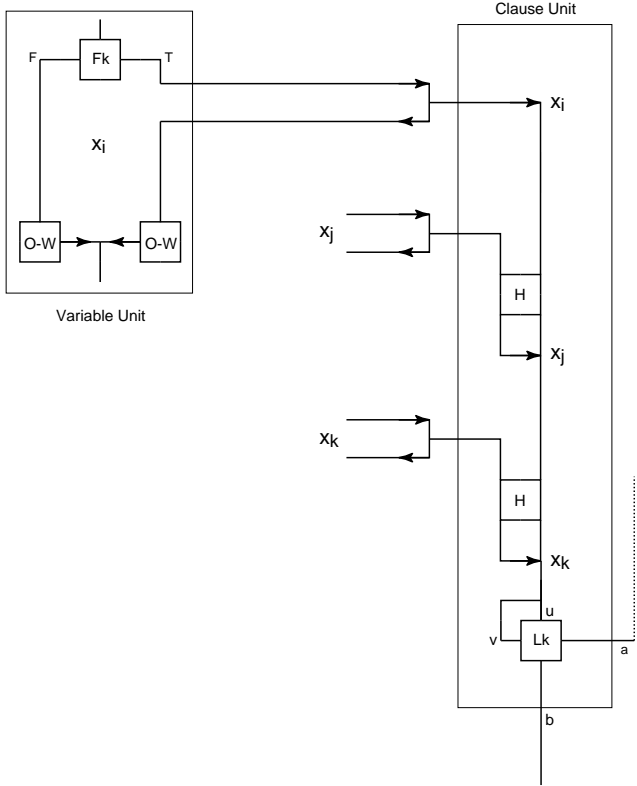}
\caption{The connections between a variable unit (left) and
a clause unit (right). 
Here $i < j < k$.
Notation: 
{\sf O-W} = One-Way;
{\sf Fk} = Fork;
{\sf Lk} = Lock;
{\sf H} = H-gadget.
}
\figlab{Clause}
\end{figure}

\subsection{H-Gadget}
The {\em H\/}-gadget, shown in
Fig.~\figref{XOR.tracks},\footnote{
	This version was suggested by Michael Hoffmann
	[personal communication, Aug. 2000].
}
would be more accurately named
a ``parallel tracks XOR''; the symbol `H' is chosen to
indicate parallel tracks with some interaction.
The following lemma summarizes its properties.

\begin{figure}[htbp]
\centering
\includegraphics[width=0.8\linewidth]{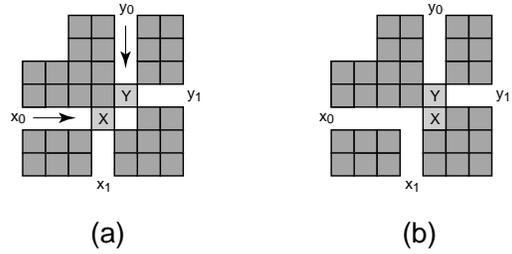}
\caption{(a) H-gadget in initial configuration; (b) after passage
through $x$-corridor.}
\figlab{XOR.tracks}
\end{figure}
\begin{lemma}
The H-gadget in its initial configuration (Fig.~\figref{XOR.tracks}a)
may be traversed from $x_0$ to $x_1$ ({\em $x$-passage\/}),
or from $y_0$ to $y_1$ ({\em $y$-passage\/}), 
but not in the reverse directions.
After $x$-passage, $y$-passage is no longer possible 
(Fig.~\figref{XOR.tracks}b),
and after $y$-passage, $x$-passage is no longer possible.
\lemlab{XOR.tracks}
\end{lemma}
\begin{pf}
Clear by inspection.
\end{pf}

\subsection{No-Reverse Gadget}
Say a gadget with distinct entrance points is {\em traversed\/}
if the robot enters at one point and exits at another.
Notice that the
One-Way gadget is ``destroyed'' by (forward) traversal, in that
subsequently it may be traversed in either direction.
We will 
need a
{\em No-Reverse\/} gadget, shown in Fig.~\figref{NR},
both to enforce the directionality of the variable-clause wires,
and later as a subcomponent of a crossover
(Section~\secref{Crossovers}).

\begin{figure}[htbp]
\centering
\includegraphics[width=0.5\linewidth]{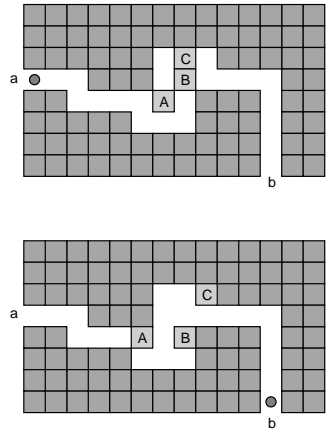}
\caption{A No-Reverse gadget before (above) and after (below) traversal.}
\figlab{NR}
\end{figure}

\begin{lemma}
The No-Reverse gadget may be traversed forward from
$a$ to $b$, but after forward traversal,
it may not be next traversed in reverse from $b$ to $a$.
\lemlab{NR}
\end{lemma}
\begin{pf}
Block $A$ must be moved leftward to leave room for $B$ to be
moved down.  The moved position of $A$ then blocks access
to $a$ from inside the gadget, preventing reversal.
(Note, however, that two forward traversals render it
an open corridor.)
\end{pf}

\subsection{The Lock}
The lock and key mechanism for PushPush used in Fig.~\figref{3D.SAT}
is straightforward, with key blocks preventing the full slide of
a necessarily pushed block.
Our Push-1
{\em Lock\/} gadget, shown in its
initial configuration in
Fig.~\figref{LD0}, is more intricate.
It has four access points, labeled $a$, $b$, $u$, and $v$.
Passage from $a$ to $b$ is blocked by
a locked
``door'' composed of blocks $A,\ldots,K$.
The ``key'' block $L$ can be accessed via $u$
and pushed to {\em unlock\/} the door, then permitting $a$-to-$b$ passage.
\begin{figure}[htbp]
\centering
\includegraphics[width=0.5\linewidth]{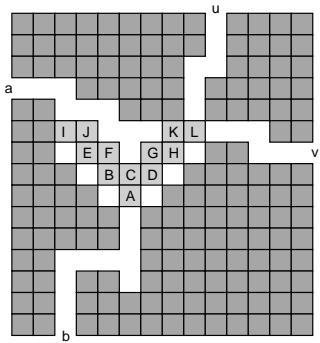}
\caption{Initial configuration of Locked Door.}
\figlab{LD0}
\end{figure}

\begin{lemma}
A Lock has the following properties:
\begin{enumerate}
\squeezelist
\item Upon first encounter, it cannot be traversed from
any of $\{ a, b, v \}$;
only passage from $u$ to $v$ is possible.
\item 
After entrance from $u$, only $v$ can be reached.
This remains true even if re-entered from $u$ later.
\item
After entrance from $u$, the state of the gadget may be altered
(unlocked) to permit later
passage from $a$ to $b$.
\item After such later $a$-to-$b$ traversal,
all of $\{ a, b, u \}$ are open to each other through
the gadget.
\end{enumerate}
\lemlab{LD}
\end{lemma}
\begin{pf}
That traversal is blocked from three of the entrance points
is clear by inspection of Fig.~\figref{LD0}.
The door is unlocked by entrance from $u$ and pushing $L$ down.
Note that from here $K$ can be pushed left (and $H$ can be pushed down, etc.),
but neither $a$ nor $b$ is accessible.

After unlocking from $u$ and entrance from $a$, a series of movements can
be made that eventually give the robot access to $A$ from above.
Start with four moves:
push $K$ right, $H$ down, $G$ right, $D$ down.
The configuration here is shown in 
Fig.~\figref{LD1}.
\begin{figure}[htbp]
\centering
\includegraphics[width=0.5\linewidth]{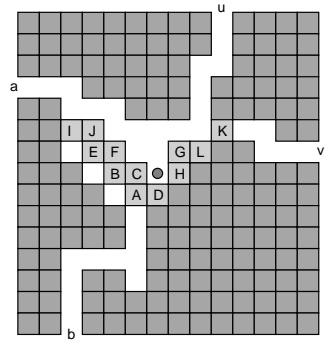}
\caption{After unlocking, partially traversed.}
\figlab{LD1}
\end{figure}
Now the ``wall'' to the left can be methodically moved
by pushing $I$ down, $J$ left, $E$ down, $F$ left,
$B$ down,
and $C$ left (or right).
Now $A$ can be pushed down the vertical corridor,
reaching a state 
(Fig.~\figref{LD2})
\begin{figure}[htbp]
\centering
\includegraphics[width=0.5\linewidth]{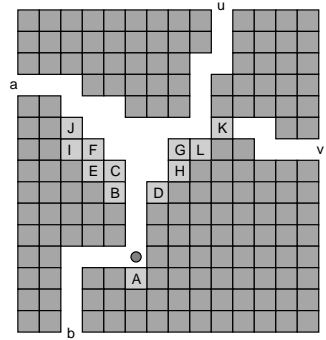}
\caption{After complete traversal.}
\figlab{LD2}
\end{figure}
where $\{a, b, u \}$ are
mutually accessible, but $v$ is cut off,
as claimed in the lemma.
\end{pf}

\subsection{Clause Unit}
The clause unit in Fig.~\figref{3D.SAT} employed one key block per literal,
which made for a simple construction.
Using the Lock as just described requires all literals to
access the same unlocking point $u$, in essence sharing 
the same key for the lock.
A naive joining of the literal lines at $u$ would permit 
several types of pernicious leakage between the lines.
Insulation can be achieved by the arrangement
shown in the schematic
in Fig.~\figref{Clause}(right).

\begin{lemma}
A Clause unit may be traversed from $a$ to $b$ only if it has been
visited from an incident literal.
Such a visit from a literal does not give access
to points $a$ or $b$; nor does it
permit leakage from one literal wire to
another.
\end{lemma}
\begin{pf}
From the $x_i$ variable unit in 
Fig.~\figref{Clause}, the robot can reach point $x_i$
in the clause,
unlock the lock via entrance $u$, returning out exit $v$
and back up to $x_i$.  From there it will be shown later
that it may only return to its variable unit along the
lower directed path.

As the robot traverses the $x_i$-$u$ path in the clause unit,
it passes through the H-gadgets, closing off later access
via $x_j$ and $x_k$ by Lemma~\lemref{XOR.tracks}.
This ensures that the lock can only be unlocked once, by
one of the literals:  whichever literal path is traversed
first necessarily closes off the other literal paths.
This prevents leakage between literals.

Once the lock is opened by one of
the literals with access to unlocking point $u$, it may be 
later traversed
from $a$ to $b$ by Lemma~\lemref{LD}(3).
\end{pf}

We now have assembled enough parts to claim that Push-1 is NP-hard in
3D, for we have designed substitutes for all components in
Fig.~\figref{3D.SAT}.
It remains to construct a crossover.

\subsection{Crossovers}
\seclab{Crossovers}
In 3D, a general crossover without leakage is
trivial, permitting passage in either direction an arbitrary
number of times.
Unfortunately it seems impossible to construct such a powerful gadget for both
Push1 and PushPush in 2D.
For our original 2D PushPush proof~\cite{ddo-ppnph2d-00}, we designed
a bidirectional crossover that could be traversed
once in each of the four directions, but was partially ``destroyed''
by each traversal, so that subsequent crossings are not possible.
This suffices for the proof, as there is never any need to
visit a Clause unit twice from the same Variable unit.
However, we were unable to mimic the functionality of our complex 
``double lock gadget''
from~\cite{ddo-ppnph2d-00} for a Push-1 robot.
Instead, 
we found it necessary to further exploit
properties of the Variable-Clause visits, and in particular, 
to enforce directionality, and
to exploit a natural sorting of the visits.
Let the Variable units be labeled $x_1, \ldots, x_n$ 
(as in Fig.~\figref{3D.SAT});
the linking of these units then ensures $x_i$ is traversed
prior to $x_j$ for $i < j$.
Our construction will arrange the wires so that
the vertical ($n$-$s$) wire at a crossover will
always be traversed prior to the horizontal ($w$-$e$) wire,
and always at most once.
We describe the crossover construction in three stages:
\begin{enumerate}
\squeezelist
\item XOR Crossover
\item Limited Unidirectional Crossover
\item Bidirectional Crossover arrangement
\end{enumerate}

\subsubsection{XOR Crossover}
The {\em XOR Crossover\/} is used in two places.
First, horizontal wires from the {\sc f}-side of a particular
Variable unit cross the vertical {\sc t}-wire 
of that unit (cf.~Fig.~\figref{3D.SAT}).
The Variable unit construction ensures that passage through the crossover
will be either via the vertical wire, or the horizontal,
but never both; so an ``exclusive-or'' crossover suffices
here.  Second, an XOR Crossover 
will be embedded inside
the more complex Limited Unidirectional Crossover described
in Section~\secref{Unidirectional}.
The XOR Crossover shown in Fig.~\figref{XOR} has these properties:
\begin{lemma}
An XOR Crossover may be traversed from $x_0$ to $x_1$
without leakage to $y_0$ or $y_1$, or from $y_0$ to $y_1$ without leakage
to $x_0$ or $x_1$.
\end{lemma}
\begin{pf}
Consider passage from $x_0$ to $x_1$.  This requires pushing block $X$
rightward, which then seals off $y_1$.  And $y_0$ is sealed off by
block $Y$.  The claim for the other direction follows from the
symmetry of the design.
\end{pf}
\begin{figure}
\centering
\includegraphics[width=0.5\linewidth]{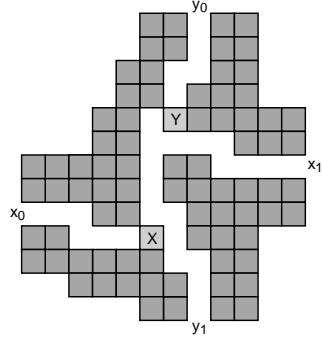}
\caption{XOR Crossover.
}
\figlab{XOR}
\end{figure}

\subsubsection{Limited Unidirectional Crossover}
\seclab{Unidirectional}
The key component to our crossover design is
what we call a {\em Limited Unidirectional Crossover} ({\sf LUC}),
whose core is shown in
Fig.~\figref{cross.all.0}.
It is limited in that it relies on the vertical being
traversed prior to the horizontal (if both are),
and unidirectional in that passage is only permitted in
one direction along the wires.  
It is also limited in that it is designed to be traversed
at most once in each direction.
Section~\secref{Bidirectional}
will extend to bidirectionality.

The four entrance/exit points are labeled $n$, $s$, $e$, $w$.
Not shown in the figure are two No-Reverse gadgets after the
$e$ and $s$ exits preventing return.
Entrances from $w$ and $n$
feed into an XOR crossover.
The $e$ entrance is
protected from entrance by a One-Way gadget, but
such protection is superfluous for the $s$ entrance.
The remainder of the design consists of two (differently oriented)
Locks (L1 and L2), and two No-Reverse gadgets (NR1 and NR2).
Its essential behavior is captured by this lemma:

\begin{lemma}
A Limited Unidirectional Crossover,
in its initial state, may not be entered from $e$ or from $s$.
It may be traversed:
\begin{enumerate}
\squeezelist
\item $w$-to-$e$
without leakage to $n$ or $s$; or
\item
$n$-to-$s$
without leakage to $w$ or $e$; or
\item
$n$-to-$s$ followed later by $w$-to-$e$ passage.
\end{enumerate}
\lemlab{LUC}
\end{lemma}
\begin{pf}
Initial entrance from $e$ is stopped by block $W_3$ in a One-Way
gadget, and entrance from $s$ is stopped by block $L_2$ of
lock L2.
We now detail the three possible traversals.
\begin{enumerate}
\item $w$-to-$e$.
If passage is through the XOR, then the only possible
leakage is to $s$ via L2.  But L2 is locked and cannot be traversed
in that direction, $b_2$ to $v_2$, by Lemma~\lemref{LD}(1).
\item $n$-to-$s$.
Passage from $n$ through the XOR necessarily unlocks L1.
From point $v_1$, there are two options: to $c_1$, through NR1,
and entrance $a_2$ of L2.  But further progress along this
route is not possible, and in fact the robot is now stuck
because of the No-Reverse unit.
The second option, to $c_2$, through NR2, brings the robot
to $u_2$, the unlocking entrance to L2.  After unlocking L2,
the robot reaches $s$.  At no point is it possible to access
$e$ or $w$, because Lemma~\lemref{LD}(1) guarantees that only
the $u_2$ to $v_2$ passage through L2 is possible.
\item $n$-to-$s$ then $w$-to-$e$.
After $n$-to-$s$ passage, both L1 and L2 are unlocked, as we just
noted.  Consider now an attempt at a $w$-to-$e$ passage.
The XOR is blocked (by $Y$) from the earlier  $n$-to-$s$ traversal.
But the robot can instead go through L1 from $a_1$ to $b_1$,
and then via $c_1$ to L2, passing through it from $a_2$ to $b_2$.
\end{enumerate}
\end{pf}
Note that the lemma makes no claims about repeated passages,
as the overall design will prevent this possibility.

\begin{figure*}[t]
\centering
\includegraphics[width=\linewidth]{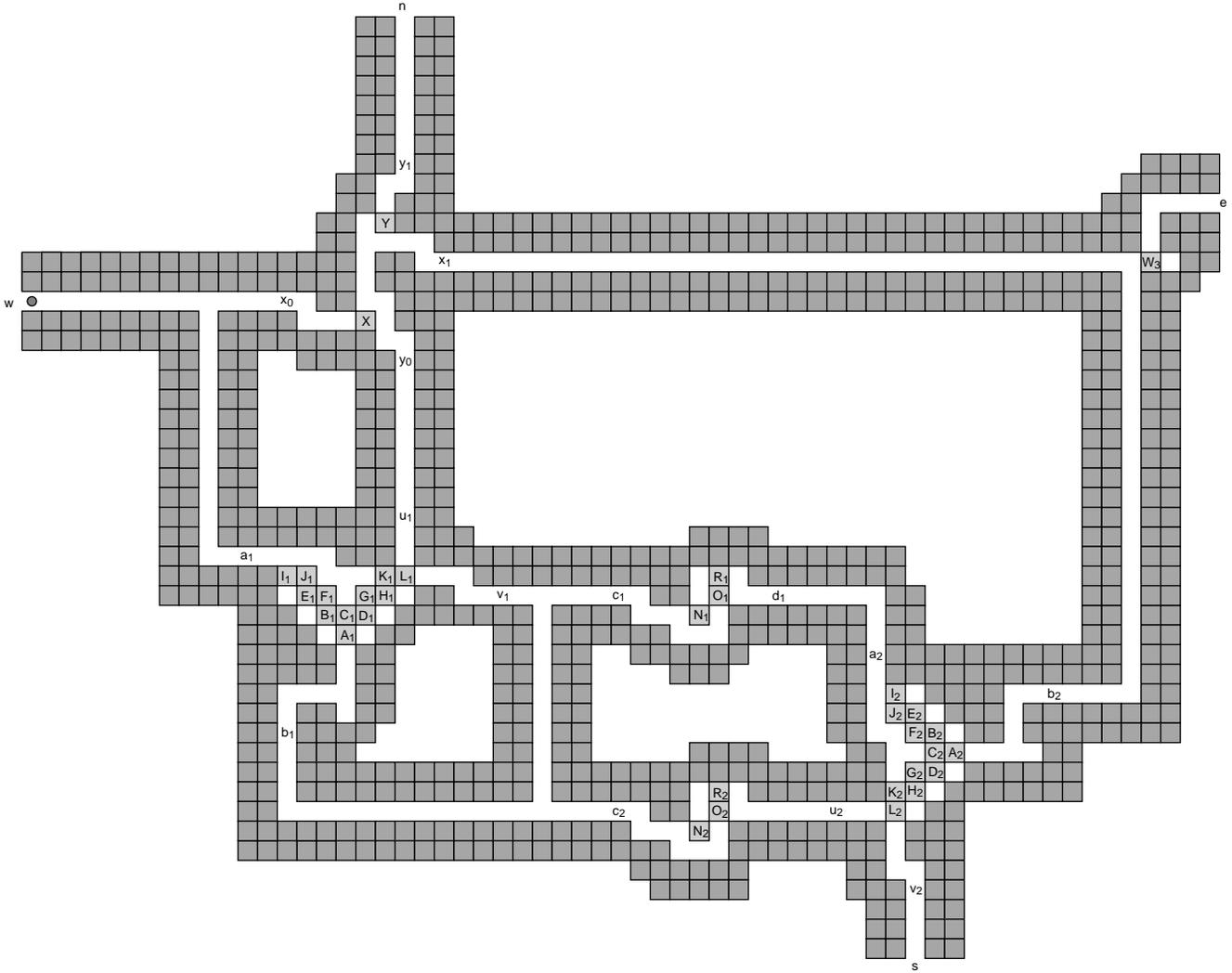}
\caption{A Limited Unidirectional Crossover. Although all blocks
are movable, only the lightly shaded blocks might be moved.
Not shown are NR gadgets beyond the $e$ and $s$ exits.}
\figlab{cross.all.0}
\end{figure*}

\subsection{Bidirectional Crossover}
\seclab{Bidirectional}
We achieve bidirectionality by arranging
Limited Unidirectional Crossovers together in the pattern shown in 
Fig.~\figref{Bicross} for any pair of literal (variable-clause) wires
that cross.
Recall from Fig.~\figref{Clause} that a literal wire is in fact
two parallel wires, one intended for moving variable-to-clause,
the other for clause-to-variable.
The directionality of these wires is enforced by the properties
and orientation of the {\sf LUC}s along it: Lemma~\lemref{LUC}
guarantees they may not be entered initially from $e$ or $s$,
and the NR gadgets at these exits ensure that reverse traversal
is not possible.
The two wires in Fig.~\figref{Bicross} are labeled
$1$ and $2$, with wire $1$ from Variable unit $x_i$ and
$2$ from Variable unit $x_j$ with $i < j$.  Thus the
$1$-wire will always be traversed first, and the crossover
exploits this; this is one sense it which it is limited.
Each Limited Unidirectional Crossover is oriented so that its ``local'' 
$n$-to-$s$ is wire $1$,
and $w$-to-$e$ wire $2$.

\begin{figure}
\centering
\includegraphics[width=0.9\linewidth]{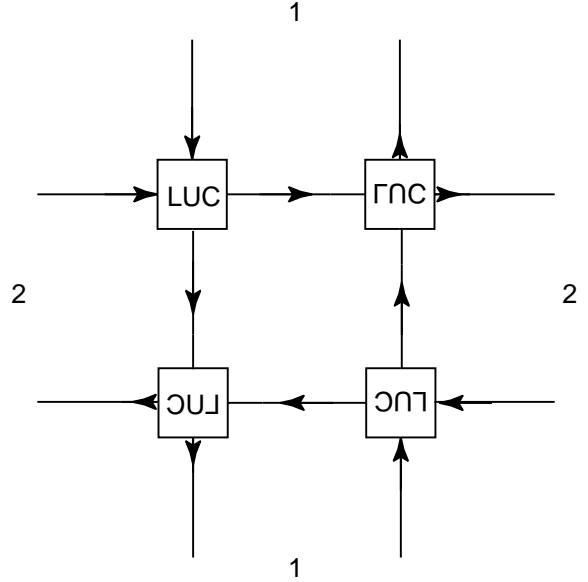}
\caption{Bidirectional Crossover.
{\sf LUC} = Limited Unidirectional Crossover 
(Fig.~\protect\figref{cross.all.0}), oriented with
$w$ left of `{\sf L}' and $n$ above `{\sf U}'.}
\figlab{Bicross}
\end{figure}

\begin{lemma}
A Bidirectional Crossover permits passage
\begin{enumerate}
\squeezelist
\item forward and back along wire $1$; or 
\item forward and back along
wire $2$; or 
\item forward and back along wire $1$ followed by 
forward and back along wire $2$.
\end{enumerate}
All these passages avoid leakage as long as each unit is
traversed at most once in each allowable direction.
\end{lemma}
\begin{pf}
The claimed properties follow directly from Lemma~\lemref{LUC}
and the design.
Consider top-to-bottom passage along the $1$-wire,
common to claims~(1) and~(3).
This takes the robot through the left two 
{\sf LUC}s,
leaving them, by Lemma~\lemref{LUC}(3), in a state to permit
later passage left-to-right (through the top left {\sf LUC})
and right-to-left (through the bottom left {\sf LUC}),
both without leakage.

Consider left-to-right passage along
the $2$-wire without prior traversal of the $1$-wire,
i.e., claim~(2) of the lemma.
The robot faces two {\sf LUC}s: the $e$-entrance
of the lower {\sf LUC} and the $w$-entrance of the upper {\sf LUC}.
By Lemma~\lemref{LUC}, the $e$-entrance is blocked, so the
robot may only pass through the upper {\sf LUC}.
Again leakage is prevented to $n$ or to $s$ through this
and the upper-right {\sf LUC} as well.  The lower {\sf LUC}s
are accessible and available for the return trip.
\end{pf}

\subsection{Overall Behavior}
\seclab{Overall}
Consider the robot making a choice of {\sc t} on variable
$x_i$.  If $x_i$ appears in some clause $C$, the robot is
forced by the design to travel down the variable-to-clause
wire, as in Fig.~\figref{Clause}.  As it crosses a
literal wire for $x_j$ with $j < i$, it crosses $w$-to-$e$;
as it crosses a
literal wire for $x_k$ with $k > i$, it crosses $n$-to-$s$.
By the design of the {\sf LUC}s, it both can do this,
and is prevented from deviating from the literal
path while doing so.  When it reaches the clause unit $C$, there are
two possibilities.  First, the clause was previously visited
along an earlier literal wire, in which case the closed H-gadgets
leave it no choice but to return along the clause-to-variable
wire without entering $C$.  (Note that then it must continue lower
along the {\sc t}-wire within the variable component: 
it cannot back up and revisit
an earlier literal wire.)
Second, the clause has not yet been visited, in which case
it has the option of unlocking the clause Lock and returning
to its variable component.  If in this case it opts not to unlock the
Lock, then only if some other literal is selected later could
the lock be successfully unlocked, permitting later passage
along the final clause-threading wire.

\begin{table*}[htbp]
\begin{center}
\begin{tabular}{| c | c | c | c | c | c | c | c | c |}
        \hline
\mbox{} & 1 & 2 & 3 & 4 & 5 & 6 & 7 & \mbox{} \\
{\em Name} &
{\em Push?}
        & {\em Blocks}
        & {\em Fixed?}
        & {\em \#}
        & {\em Path?}
        & {\em Dim}
        & {\em Sliding}
        & {\em Complexity}
        \\ \hline \hline
\mbox{} &
pull
        & L
        & fixed
        & $k$
        & path
        & 2D
        & min
        & NP-hard \cite{w-mppmo-91}
        \\ \hline
\mbox{} &
push
        & unit
        & fixed
        & $k$
        & path
        & 2D
        & min
        & NP-hard \cite{do-mpams-92}
        \\ \hline
Sokoban &
push
        & unit
        & movable
        & $1$
        & storage
        & 2D
        & min
        & NP-hard \cite{dz-sompp-95}
        \\ \hline
Sokoban &
push
        & unit
        & movable
        & $1$
        & storage
        & 2D
        & min
        & PSPACE \cite{c-spc-99}
        \\ \hline
PushPush3D &
push
        & unit
        & movable
        & $1$
        & path
        & 3D
        & {\em max}
        & NP-hard \cite{ppnph-os-99}
        \\ \hline
\mbox{} &
push
        & unit
        & movable
        & $1$
        & storage
        & 2D
        & {\em max}
        & NP-hard \cite{ppnph-os-99}
        \\ \hline \hline

{\bf PushPush2D} &
{\bf push}
        & {\bf unit}
        & {\bf movable}
        & {\bf 1}
        & {\bf path}
        & {\bf 2D}
        & {\bf {\em max}}
        & {\bf NP-hard}
        \\ \hline \hline

{\bf Push-$1$} &
{\bf push}
        & {\bf unit}
        & {\bf movable}
        & {\bf 1}
        & {\bf path}
        & {\bf 2D}
        & {\bf {\em min}}
        & {\bf NP-hard}
        \\ \hline \hline
Push-$*$&
push
        & unit
        & movable
        & $k$
        & path
        & 2D
        & min
        & NP-hard \cite{h-p*nph-00}
        \\ \hline
Push-$1X$ &
push
        & unit
        & movable
        & $1$
        & noncrossing
        & 2D
        & min
        & open$^3$
        \\
\mbox{} &
\mbox{}
        & \mbox{}
        & \mbox{}
        & \mbox{}
        & path
        & \mbox{}
        & \mbox{}
        & \mbox{}
        \\ \hline
\end{tabular}
\end{center}
\caption{Pushing block problems.}
\vspace{2ex}
\tablab{results}
\end{table*}

\subsection{Main Theorem}
Now the conclusion that Push-1 is NP-hard follows immediately,
for we have successfully constructed all the components necessary.
The overall design continues to follow Fig.~\figref{3D.SAT}, with
additional turns in the wires to arrange all crossovers to 
respect the ordering of the wires crossed.
Finally, a review of each constituent of the construction
shows that all retain their essential properties even if
the robot has PushPush powers.
We may conclude:
\begin{theorem}
PushPush and Push-1 are both NP-hard in 2D.
\theolab{pp.p1}
\end{theorem}

We leave it open whether 
Theorem~\theoref{pp.p1} can be
strengthened in either direction:
either by proving either problem is in NP,
in which case it is NP-complete,
or by showing that either is PSPACE-complete.

\section{Summary}
We conclude by summarizing 
in Table~\tabref{results}
previous work according to
the classification scheme offered in Section~\secref{Introduction},
and comparing it to recent work.
The first six lines show previous results,
including 
the results from~\cite{ppnph-os-99}.
(The 2D storage result is, incidentally, not difficult.)
The two boldface lines of the table 
are the results of this paper.

The penultimate line of the table describes a recent result
by Hoffmann~\cite{h-p*nph-00}:
``{\em Push-$*$\/}'' is NP-hard, where
all blocks are movable and the robot can push an
arbitrary number of blocks, sliding the minimal amount.
This settles an open problem from~\cite{do-mpams-92}.

Finally, the last line of the table suggests a new open
problem with the same characteristics as Push-1, but
with the added stipulation that the robot never revisit
a square it previously occupied.
It is easy to see that this new problem, which
we dub {\em Push-$1X$}, is in NP, which already places it
on a different footing than all other problems.
Perhaps Push-$1X$ (or some variation thereof) is in P?\footnote{
	Note added Aug. 2000:  A new proof by
	M. Hoffmann and E. Demaine [forthcoming] shows
	that even Push-$1X$ is NP-hard.
}

\section*{Acknowledgments}

We thank Therese Biedl for helpful discussions.
The third author acknowledges many insights from meetings of the
Smith Problem Solving Group.

Finally, we are grateful to Thomas Shermer 
for uncovering an error
in our proof, and helping to fix it
[personal communication, Aug. 2000].
In particular our earlier
version\footnote{
	Presented at the {\em 12th Canad. Conf. Comput. Geom.}, Aug. 2000.
} 
permitted some {\sf LUC}s to be retraversed, which led to
leakage that undermined the reduction.

\bibliographystyle{alpha}
\bibliography{/home1/orourke/bib/geom/geom}
\end{document}